\documentclass{aa}
\usepackage{txfonts}
\usepackage{longtable}
\usepackage{graphicx}
\newcommand{\feh} {\mbox{\rm [Fe/H]}}

\newcommand{\nafe} {\mbox{\rm [Na/Fe]}}

\newcommand{\nife} {\mbox{\rm [Ni/Fe]}}
\newcommand{\cufe} {\mbox{\rm [Cu/Fe]}}
\newcommand{\znfe} {\mbox{\rm [Zn/Fe]}}

\newcommand{\alphafe} {\mbox{\rm [$\alpha$/Fe]}}
\newcommand{\teff}  {\mbox{$T_{\rm eff}$}}

\newcommand{\logg}  {\mbox{{\rm log}\,$g$}}

\newcommand{\LiI} {\ion{Li}{i}}

\newcommand{\NaI} {\ion{Na}{i}}

\newcommand{\FeI} {\ion{Fe}{i}}
\newcommand{\FeII} {\ion{Fe}{ii}}

\newcommand{\VK}{\mbox{($V\!-\!K)$}}

\newcommand{\by}{\mbox{($b\!-\!y)$}}

\begin{document}

\title{Two distinct halo populations in the solar neighborhood}

\subtitle{IV. Lithium abundances
\thanks{Based on observations made with the Nordic Optical Telescope
on La Palma, and on data from the European Southern Observatory
ESO/ST-ECF Science Archive Facility (programs 65.L-0507,
67.D-0086, 67.D-0439, 68.D-0094, 68.B-0475, 69.D-0679,
70.D-0474, 71.B-0529, 72.B-0585, 76.B-0133 and  77.B-0507).}
\fnmsep\thanks{Table 1 is provided as online material
and is available in electronic form at {\tt http://www.aanda.org}.}}

\author{P. E. Nissen \inst{1} \and W. J. Schuster \inst{2}}

%\offprints{P.E.~Nissen}

\institute{
Department of Physics and Astronomy, University of Aarhus, DK--8000
Aarhus C, Denmark.
\email{pen@phys.au.dk}
\and Observatorio Astron\'{o}mico Nacional, Universidad Nacional Aut\'{o}noma
de M\'{e}xico, Apartado Postal 877, C.P. 22800 Ensenada, B.C., M\'{e}xico.
\email{schuster@astrosen.unam.mx}}

\date{Received 4 April 2012 / Accepted 13 May 2012}

\abstract
% context heading (optional)
{A previous study of F and G main-sequence stars  
in the solar neighborhood has revealed the existence of
two distinct halo populations with a clear separation in
[$\alpha$/Fe] for the metallicity range $-1.4 <$ \feh $< -0.7$. 
Taking into account the kinematics and ages of the stars, some 
Galactic formation models suggest that the `high-alpha' halo  stars were formed 
{\em in situ}, whereas the `low-alpha' stars have been accreted from
satellite galaxies.} 
% aims heading (mandatory)
{In this paper we investigate if there is a systematic difference
in the lithium abundances of stars belonging to the high- and
low-alpha halo populations.}
% methods heading (mandatory)
{Equivalent widths of the \LiI\ 6707.8\,\AA\ resonance line
are measured from high resolution 
VLT/UVES and NOT/FIES spectra and used to derive
Li abundances on the basis of MARCS model atmospheres.
Furthermore, masses of the stars are determined from the
log\,\teff -- \logg\ diagram by interpolating between evolutionary
tracks based on Yonsei-Yale models.}
% results heading (mandatory)
{There is no significant systematic difference in the lithium abundances of
high- and low-alpha stars. For the large majority of stars with masses
$0.7 < M/M_{\odot} < 0.9$ 
and heavy-element mass fractions $0.001 \la Z < 0.006$, the lithium 
abundance is well fitted by a relation 
$A$(Li) = $a_{0}$ + $a_{1} \, M$ + $a_{2} \, Z$ + $a_{3} \, M \, Z$,
where $a_{0}$, $a_{1}$, $a_{2}$, and $a_{3}$ are constants.
Extrapolating this relation to $Z = 0$ leads to a lithium abundance
close to the primordial value predicted from standard 
Big Bang nucleosynthesis calculations and the WMAP baryon density.
The relation, however, does not apply to stars with metallicities
below $\feh \simeq -1.5$.}
% conclusions heading (optional)
{We suggest that metal-rich halo stars were formed
with a lithium abundance close to the primordial value, and that lithium
in their atmospheres has been depleted in time with an approximately linear
dependence on stellar mass and $Z$. The lack of a systematic
difference in the Li abundances of high- and low-alpha stars
indicates that an environmental effect is not important for the destruction
of lithium.}


\titlerunning{Lithium abundances in two distinct halo populations}
\authorrunning{Nissen \& Schuster}

\maketitle

\newpage

\section{Introduction}
\label{sect:introduction} 
The discovery by Spite \& Spite (\cite{spite82a}, \cite{spite82b})
of a uniform lithium abundance in halo dwarf and subgiant stars with
effective temperatures
$ \teff \ga 5700$\,K has led to important advances in the fields
of stellar physics and Big Bang nucleosynthesis.
Recent works on lithium abundances in very metal-poor stars raise, 
however, severe  problems in our understanding of how lithium was
synthesized in the Big Bang and subsequently destroyed in stars. 

The work of Spite \& Spite (\cite{spite82b})
showed that eight halo stars with $5700 < \teff < 6300$\,K and 
$-2.4 < \feh < -1.4$ have an average lithium abundance
$A$(Li) = log($N_{\rm Li} / N_{\rm H})$ + 12 = 2.06 with  a rms dispersion
of 0.10\,dex and no significant trends as a function of \teff\ or \feh .
Later, this so-called `Spite plateau' has been confirmed for larger samples
of halo stars (e.g. Molaro et al. \cite{molaro95}; Spite et al. 
\cite{spite96}). More recently, a very precise determination
of lithium isotopic abundances by Asplund et al. (\cite{asplund06}) 
showed $A$(Li) to be constant at 2.23 in the metallicity 
range $-2.5 < \feh < -1.4$ with a dispersion of 0.03\,dex only.
A revised \teff -scale explains
the higher Li-plateau abundance as compared to Spite \& Spite
(\cite{spite82b}). Furthermore, Asplund et al. find that the
bulk of lithium (i.e. more than $\sim \! 95$\%) consists of the $^7$Li
isotope.

Although the data of Asplund et al. (\cite{asplund06}) support the
existence of a Li-plateau from $-2.5$ to $-1.4$ in \feh , there is a 
a hint of a decline of $A$(Li) for five stars with  $\feh < -2.5$.
This is in agreement with an earlier investigation by Ryan et al. 
(\cite{ryan99}), who found a dependence of $A$(Li) on metallicity for
a sample of 23 very metal-poor halo stars; $A$(Li) declines from 2.15 at
$\feh = -2.5$ to 2.03 at $\feh = -3.5$.  Several recent 
works (Boesgaard et al. \cite{boesgaard05},
Bonifacio et al. \cite{bonifacio07}; Aoki et. al. \cite{aoki09};
Sbordone et al. \cite{sbordone10}) also suggest a `melt-down' of the Spite plateau 
by finding a decrease and/or increased scatter in $A$(Li) as a function
of decreasing metallicity below $\feh \simeq -2.5$. In this connection, the work of
Mel\'{e}ndez et al. (\cite{melendez10}) is of particular
interest. For a sample of 88 dwarf halo stars with $-3.5 < \feh < -1.0$
having precise  measurements of the equivalent width of the 6707.8\,\AA\
\LiI\ line, they find a break in $A$(Li) at $\feh \simeq -2.5$. 
Above this metallicity, the stars distribute around  a plateau at
$A$(Li) = 2.27, whereas stars in the metallicity range
$-3.5 < \feh < -2.5$ have an average lithium abundance of $A$(Li) = 2.18.
This two-step distribution of $A$(Li) is particularly striking if
one selects stars according to a metallicity dependent limit in
\teff\ (see Fig. 3 in Mel\'{e}ndez et al. \cite{melendez10}). 
Apparently, the scatter in $A$(Li) below $\feh \simeq -2.5$, as 
found in some of the works cited above, is due to an
increased degree of lithium depletion in the lower-mass stars.

The Spite plateau was originally interpreted as evidence of the primordial
lithium abundance and used to constrain the 
baryon density in the Big Bang from nucleosynthesis
calculations. Adopting the WMAP value of the cosmic baryon density  
(Komatsu et al. \cite{komatsu11}), standard Big Bang nucleosynthesis 
predicts, however, a primordial lithium abundance, $A$($^7$Li) = $2.72 \pm 0.06$
(Cyburt et al. \cite{cyburt08}; Coc et al. \cite{coc12}), i.e. 
about a factor of three higher than the abundance corresponding to the Li-plateau.

A possible explanation of this so-called `lithium problem' could be that
lithium in the atmospheres of stars on the Spite plateau is
depleted by gravitational settling and/or by mixing with gas from layers with such a 
high temperature ($T \ga 2.5 \, 10^6$\,K) that Li is destroyed by
reactions with protons. Stellar models including
rotational induced mixing (Pinsonneault et al. \cite{pinsonneault99}; 
Piau \cite{piau08}),
atomic diffusion (Salaris \& Weiss \cite{salaris01}; Richard et al.
\cite{richard02}) and internal gravity waves (Talon \& Charbonnel
\cite{talon04}) predict a significant Li depletion, but they have difficulties
in explaining a depletion of 0.5\,dex and a Li-plateau with very
little scatter in $A$(Li). The models of Richard et al. (\cite{richard05}), 
which include atomic and turbulent diffusion, are more successful,
when a free parameter describing the degree of turbulent mixing 
is optimized. It remains, however, to be seen if such
models can explain the decline in $A$(Li) for $\feh \la -2.5$;
Richard et al. argue in fact that Li depletion in stars 
on the main sequence does not depend on metallicity.

In addition to `in situ' stellar depletion of lithium,  
Piau et al. (\cite{piau06}) suggest that the 
Li abundance of gas in the Galactic halo may have been
reduced by nuclear burning in massive, 
zero-metallicity (Population III) stars before low-mass stars 
on the Spite plateau formed. The decline and/or increased scatter 
of $A$(Li) at low metallicity, could then be due to incomplete
mixing of gas and preferential star formation in the vicinity of supernovae
when $\feh \la -2.5$. In order to obtain a significant reduction
of the primordial lithium abundance, say 0.3\,dex, about 50\,\%
of the gas in the early halo has to pass through Pop III stars.
If these stars have masses in the range 10 - 40 $M_{\odot}$
and current nucleosynthesis models are correct, they would, however,
produce an excessive amount of CNO elements compared to the observed abundances
in metal-poor halo stars (Prantzos \cite{prantzos07}). Very
massive stars collapsing as black holes would be better candidates
provided that they eject their Li-depleted envelope, but not
the metal-rich core.

A more radical solution of the lithium problem may be found if 
decaying supersymmetric particles play a role in Big Bang
nucleosynthesis (BBN). For the right combination of the
properties of such putative particles, the abundance of 
lithium can be reduced to a level that agrees with $A$(Li) in stars on
the Spite plateau  (see review by Fields \cite{fields11}) 
without spoiling the agreement between the deuterium abundance
calculated from standard BBN  and that measured in quasar absorption line systems
(Pettini et al. \cite{pettini08}). However, non-standard BBN cannot
explain the dependence of $A$(Li) on metallicity below 
$\feh \sim -2.5$, so some stellar depletion or Galactic astration of 
lithium must be present.

It is clear from this discussion that the lithium problem is
far from being solved and that new observations and further theoretical
works in the fields of stellar physics, Galactic evolution 
and primordial nucleosynthesis are needed. Regarding new observational
data, we note that recent efforts have concentrated on    
lithium abundances in very metal-poor stars, whereas the
metal-rich end of the Spite plateau has been somewhat neglected. 
In the present paper, we try to remedy this by making a study of
Li abundances in a sample of halo and thick-disk main-sequence stars  
with $-1.6 < \feh < -0.4$ for which we have previously
determined precise abundances of a number of elements from Na to Ba
(Nissen \& Schuster \cite{nissen10}, \cite{nissen11}; hereafter
Papers I and II). These abundances have revealed the existence of
two distinct halo populations in the solar neighborhood  with a clear separation in
[$\alpha$/Fe] \footnote{$\alpha$ refers to the average
abundance of Mg, Si, Ca, and Ti}, \nafe , \nife , \cufe , and \znfe .
The kinematics and ages (Schuster et al. {\cite{schuster12}, Paper III) of the stars
suggest that the high-alpha halo  stars were formed
{\em in situ} in the inner regions of the Galaxy, whereas the low-alpha 
stars were accreted from satellite galaxies. Hence, we have a unique
possibility to see if lithium abundances of stars formed in different
environments are the same, which is of particular interest in connection with
the suggestion of Piau et al. (\cite{piau06})  of lithium destruction in an early
generation of massive stars. Furthermore, the sample of stars allows us
to investigate trends of $A$(Li) as a function of stellar mass and
heavy-element abundance, which may be used to constrain theories
of lithium depletion in stars.

In Sect. 2, we describe how the lithium
abundance $A$(Li), stellar mass $M$, and total heavy-element mass fraction $Z$
are determined and check the errors of these parameters 
by comparing with Mel\'{e}ndez et al. (\cite{melendez10})
for 24 stars in common. In Sect. 3, the dependence of $A$(Li)
on $M$ and $Z$ is investigated. In Sect. 5, the results are discussed 
and some concluding remarks are given in Sect. 6.

\section{Determination of $A$(Li), $M$, and $Z$}
\label{sect:parameters}

\subsection{Lithium abundances}
\label{sect:lithium}
The abundance of Li is determined from the equivalent width (EW)
of the \LiI\ $\lambda 6707.8$ resonance line as measured in
high-resolution spectra obtained with UVES at the ESO/VLT and FIES
at the Nordic Optical Telescope. For a more detailed description
of these spectra we refer to Paper I.
The signal-to-noise ($S/N$) of the UVES spectra is very
high, i. e. ranging from about 200 to 600,
but some of the UVES spectra suffer from a small residual fringing
at the wavelength of the \LiI\ line.  The FIES spectra have
lower $S/N$ (140 to 200) but are free from fringing. 
The typical 1-sigma error of an EW measurement is estimated to be
$\pm 1$\,m\AA . This is confirmed by comparing data for
six stars observed with both UVES and FIES (see Paper I, Table 2).
The mean  difference (FIES $-$ UVES) is 0.2\,m\AA\ with
a rms deviation of 1.1\,m\AA. There is also very good agreement
with Mel\'{e}ndez et al. (\cite{melendez10}); for 24 stars in common,
the mean EW difference is 0.02\,m\AA\ and the rms deviation is 1.0\,m\AA.

The measured equivalent widths are given in Table \ref{table:one}.
For 25 stars, the \LiI\ line could not be detected. Instead,
two-sigma upper limits, as estimated from the $S/N$ of the spectrum,
are given. Furthermore, one star in Paper I (HD\,17820)
is excluded, because the \LiI\ line is strongly disturbed  by a 
cosmic ray hit.

For each star a model atmosphere has been obtained from the 
MARCS grid (Gustafsson et al. \cite{gustafsson08}) by interpolating 
to the \teff , \logg , \feh , 
and \alphafe\ values of the star.
The Uppsala program BSYN is used to calculate
the profile and equivalent width of the \LiI\ $\lambda 6707.8$ line 
as a function of $A$(Li) assuming local thermodynamic equilibrium (LTE).
Interpolation to the observed EW-value then yields $A$(Li). 

The wavelengths and $gf$-values of the 
hyperfine components of the \LiI\ $\lambda 6707.8$ line  are adopted 
from Sansonetti et al. 
(\cite{sansonetti95}) and Yan \& Drake (\cite{yan95});
see Smith et al. (\cite{smith98}, Table 3). $^6$Li components 
are not included. If the isotopic $^6$Li/$^7$Li ratio is on the order of 5\%,
as found by Asplund et al. (\cite{asplund06}) for some stars, the
profile of the Li line will be slightly broader than calculated, but the
EW is practically the same for a given total lithium abundance, 
because the Li line is far from being saturated. In practice, we are
therefore determining the total lithium abundance. 
With the resolutions of the spectrographs
applied, $R = 60\,000$ for UVES and $R = 40\,000$ for FIES, it is
not possible to detect $^6$Li at a level of 5\,\%, but from the 
relation between the center-of-gravity (cog) wavelength of the \LiI\ line and 
the $^6$Li fraction (Smith et al. \cite{smith98}, Eq. 1),
the measured cog-wavelengths indicate that
the $^6$Li/$^7$Li ratio is less than about 10\%. 

Effective temperatures and surface gravities are determined spectroscopically,
i.e. \teff\ from the excitation balance of weak \FeI\ lines and \logg\
from the ratio of iron abundances derived from \FeI\ and \FeII\ lines.
As described in Paper I, the zero points for these parameters
are set by two nearby, unreddened `standard' stars,
\object{HD\,22879} and \object{HD\,76932}. 
Their effective temperatures were determined from the color indices
\by\ and \VK\ using the calibrations of
Ram\'{\i}rez \& Mel\'{e}ndez (\cite{ramirez05}). 
The recent more accurate calibrations by Casagrande et al.
(\cite{casagrande10}) shows, however, a systematic offset of 
about +100\,K in \teff\ relative to the
Ram\'{\i}rez \& Mel\'endez values for stars with $\feh > -2.0$ and
$4800 < \teff < 6200$\,K.  We have, therefore, increased the \teff\ values
given in Paper I by 100\,K. As discussed in Paper III, this leads to a  
systematic correction of +0.03\,dex of \logg\ and corrections
of \feh\ ranging from about $-0.03$\,dex at
\teff \,= 5400\,K to +0.01\,dex at \teff \,= 6100\,K, whereas
the correction of \alphafe\ is approximately constant at $-0.01$\,dex.
The +100\,K correction of \teff\ changes the derived lithium
abundances by about +0.08\,dex, whereas the corresponding changes
of \logg , \feh , and \alphafe\ have no significant effect on $A$(Li).
The values of \logg , \feh , and \alphafe\ given in Paper I have,
therefore, been adopted.

The derived LTE lithium abundances are given in Table \ref{table:one}
together with the adopted model atmosphere parameters, \teff , \logg ,
\feh , and \alphafe . The table also lists non-LTE 
lithium abundances based on the NLTE\,$-$\,LTE corrections calculated
by Lind et al. (\cite{lind09}) for 1D MARCS models and collisional
cross sections from Barklem et al. (\cite{barklem03}).
These corrections depend mainly 
on the effective temperature and range from about +0.04\,dex at 
\teff \,= 5400\,K to $-0.06$\,dex at \teff \,= 6100\,K. 3D non-LTE 
calculations are not available for the metallicities of our
stars, but for a sample of very metal-poor stars
Sbordone et al. (\cite{sbordone10}) find only small differences in $A$(Li)
($\pm 0.02$\,dex), when 1D LTE and 3D NLTE analyses are compared.

The error of $A$(Li) is mainly caused by the error of \teff .      
In Paper III, the one-sigma error of the differential \teff\
values was estimated to be on the order of $\pm 30$\,K from a 
comparison of our spectroscopic temperatures with values determined
from \by\ and \VK\ for a sample 44 stars that are unreddened according
to the absence of interstellar \NaI \,D lines. The corresponding error of
$A$(Li) is 0.025 dex. Adding the small error arising from
the EW measurements, we arrive at a statistical error of $\pm 0.03$\,dex
in $A$(Li) if the equivalent width of the \LiI\ $\lambda 6707.8$ line
is greater than about 20\,m\AA .

\begin{figure}
\resizebox{\hsize}{!}{\includegraphics{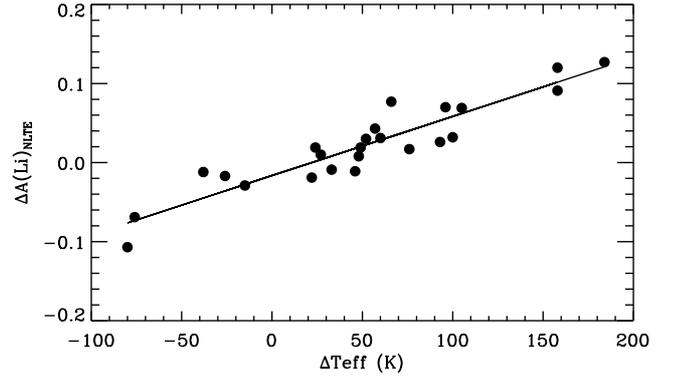}}
\caption{Differences between  Li abundances in  
Mel\'{e}ndez et al. (2012, in preparation) and in this paper
versus the corresponding differences in \teff .}
\label{fig:delALi}
\end{figure}

To check the accuracy of our Li abundances, we have compared with
Mel\'{e}ndez et al. (\cite{melendez10}) for 24 stars in common 
using a slightly 
revised version of their Table 1 (Mel\'{e}ndez et al. 2012, in preparation).
MARCS models are also used to derive their abundances, but \teff\
is determined with the infrared flux method (IRFM) as
implemented by  Casagrande et al.  (\cite{casagrande10}) and with the use of 
interstellar \NaI \,D lines to estimate the reddening. The mean 
difference (Mel\'{e}ndez -- this paper) is  $<\!\!\Delta A({\rm Li})\!\!>\, = 0.022$ 
with a scatter of $\sigma = 0.054$\,dex, which corresponds quite well to the estimated 
errors of $A$(Li), i.e. $\pm 0.035$ in Mel\'{e}ndez et al. (\cite{melendez10}) 
and $\pm 0.030$\,dex in this paper. 

As seen from  Fig. \ref{fig:delALi},
the differences in $A$(Li) are closely correlated with the corresponding
differences in \teff ; the scatter around the regression line
($\Delta A$(Li) = $-0.016 + 0.00075 \, \Delta \teff$) 
is only 0.021\,dex.  This shows that the differences in $A$(Li) between
Mel\'{e}ndez et al.  and this paper
are mainly caused by differences in \teff\  with only a small contribution 
coming from the differences in the measured equivalent widths of the   
\LiI\ $\lambda 6707.8$ line. 

The mean deviation in effective temperature
(Mel\'{e}ndez -- this paper) is  $<\!\!\Delta \teff \!\!>\, = 51$\,K with 
$\sigma = 66$\,K. 
A comparison with Casagrande et al.  (\cite{casagrande10})
for 43 stars in common results, however, in a smaller mean difference
(Casagrande -- this paper) $<\!\!\Delta \teff \!\!>\, = 16$\,K with
$\sigma = 61$\,K. Furthermore, a comparison with the IRFM temperatures of
Gonz\'{a}lez Hern\'{a}ndez \& Bonifacio (\cite{gonzalez09}) for 45 stars in common gives
a mean difference (Gonzalez -- this paper) $<\!\!\Delta \teff \!\!>\, = -16$\,K
with $\sigma = 81$\,K. Altogether, we estimate that the 
uncertainty of the \teff -scale is on the order of $\pm 50$\,K. The corresponding
error of $A$(Li) is about 0.04\,dex.

\onltab{1}{
\clearpage \onecolumn
\begin{longtable}{lcccccccrrrc}
\caption{\label{table:one}Stellar parameters and lithium abundances derived from the
equivalent width (EW) of the \LiI\ 6707.8\,\AA\ line.} \\
\noalign{\smallskip}
\hline\hline
\noalign{\smallskip}
  ID & Pop.$^{\rm a}$ & \teff  & \logg  & \feh  & \alphafe  & $Z$  & $M/M_{\odot}$ &   EW  & 
	 $A$(Li) & $A$(Li) & Note$^{\rm b}$ \\ 
     &                &   [K]  &        &       &           &      &               & [m\AA ] & 
    LTE  &  NLTE   &                \\ 
\noalign{\smallskip}
\hline
\noalign{\smallskip}
\noalign{\smallskip}
\endfirsthead
\caption{continued} \\
\hline\hline
\noalign{\smallskip}
  ID & Pop.$^{\rm a}$ & \teff  & \logg  & \feh  & \alphafe  & $Z$  & $M/M_{\odot}$ &   EW  & 
 $A$(Li) & $A$(Li) & Note$^{\rm b}$ \\ 
    &                &   [K]  &        &       &           &      &               & [m\AA ] & 
    LTE  &  NLTE   &                \\ 
\noalign{\smallskip}
\hline
\noalign{\smallskip}
\noalign{\smallskip}
\endhead
\hline
\endfoot
BD$-$21 3420 &   C &   5908 &   4.26 & $ -1.13$ &   0.31 &   0.0024 &    0.77   & $   21.2$ & $    1.95$ & $    1.92$ & \\
CD$-$33 3337 &   C &   6079 &   3.86 & $ -1.36$ &   0.30 &   0.0014 &    0.85   & $   39.2$ & $    2.37$ & $    2.32$ & \\
CD$-$43 6810 &   A &   6045 &   4.26 & $ -0.43$ &   0.23 &   0.0105 &    0.99   & $<   1.0$ & $<   0.72$ & $<   0.70$ & \\
CD$-$45 3283 &   B &   5697 &   4.55 & $ -0.91$ &   0.12 &   0.0029 &    0.79   & $<   1.0$ & $<   0.43$ & $<   0.43$ & \\
CD$-$51 4628 &   B &   6253 &   4.31 & $ -1.30$ &   0.22 &   0.0014 &    0.85   & $   29.7$ & $    2.36$ & $    2.29$ & \\
CD$-$57 1633 &   B &   5973 &   4.28 & $ -0.90$ &   0.07 &   0.0028 &    0.81   & $   30.5$ & $    2.19$ & $    2.15$ & \\
CD$-$61 282  &   B &   5859 &   4.31 & $ -1.23$ &   0.22 &   0.0016 &    0.73   & $   36.7$ & $    2.18$ & $    2.14$ & \\
G05-19       &   B &   5954 &   4.26 & $ -1.18$ &   0.19 &   0.0017 &    0.76   & $   32.7$ & $    2.20$ & $    2.15$ & \\
G05-36       &   A &   6113 &   4.23 & $ -1.23$ &   0.35 &   0.0020 &    0.83   & $   41.6$ & $    2.43$ & $    2.37$ & \\
G05-40       &   A &   5895 &   4.17 & $ -0.81$ &   0.31 &   0.0049 &    0.83   & $   13.5$ & $    1.74$ & $    1.72$ & \\
G13-38       &   A &   5363 &   4.54 & $ -0.88$ &   0.32 &   0.0044 &    0.73   & $<   3.0$ & $<   0.58$ & $<   0.61$ & \\
G15-23       &   A &   5397 &   4.57 & $ -1.10$ &   0.34 &   0.0028 &    0.71   & $<   3.0$ & $<   0.61$ & $<   0.63$ & \\
G16-20       &   B &   5725 &   3.64 & $ -1.42$ &   0.26 &   0.0011 &    0.85   & $   38.8$ & $    2.10$ & $    2.07$ & \\
G18-28       &   A &   5472 &   4.41 & $ -0.83$ &   0.31 &   0.0048 &    0.72   & $<   2.0$ & $<   0.51$ & $<   0.53$ & SB1 \\
G18-39       &   A &   6140 &   4.21 & $ -1.39$ &   0.34 &   0.0014 &    0.80   & $   30.6$ & $    2.29$ & $    2.23$ & \\
G20-15       &   B &   6127 &   4.32 & $ -1.49$ &   0.24 &   0.0010 &    0.78   & $   34.1$ & $    2.33$ & $    2.27$ & \\
G21-22       &   B &   6001 &   4.24 & $ -1.09$ &   0.09 &   0.0018 &    0.78   & $   37.9$ & $    2.31$ & $    2.25$ & \\
G24-13       &   A &   5773 &   4.31 & $ -0.72$ &   0.29 &   0.0059 &    0.82   & $<   3.0$ & $<   0.97$ & $<   0.97$ & \\
G24-25       &   A &   5928 &   3.86 & $ -1.40$ &   0.35 &   0.0014 &    0.81   & $<   2.0$ & $<   0.90$ & $<   0.87$ & SB1 \\
G31-55       &   A &   5738 &   4.30 & $ -1.10$ &   0.29 &   0.0025 &    0.72   & $   10.9$ & $    1.51$ & $    1.49$ & \\
G46-31       &   B &   6001 &   4.23 & $ -0.83$ &   0.15 &   0.0037 &    0.83   & $   26.1$ & $    2.14$ & $    2.10$ & SB1 \\
G49-19       &   A &   5872 &   4.25 & $ -0.55$ &   0.27 &   0.0085 &    0.89   & $    4.5$ & $    1.23$ & $    1.23$ & SB1 \\
G53-41       &   B &   5959 &   4.27 & $ -1.20$ &   0.23 &   0.0018 &    0.76   & $   27.2$ & $    2.11$ & $    2.06$ & \\
G56-30       &   B &   5930 &   4.26 & $ -0.89$ &   0.11 &   0.0030 &    0.80   & $   29.5$ & $    2.14$ & $    2.10$ & \\
G56-36       &   B &   6033 &   4.28 & $ -0.94$ &   0.20 &   0.0032 &    0.84   & $   24.1$ & $    2.12$ & $    2.07$ & \\
G57-07       &   A &   5776 &   4.25 & $ -0.47$ &   0.31 &   0.0107 &    0.90   & $<   2.0$ & $<   0.80$ & $<   0.81$ & \\
G63-26       &   A &   6143 &   4.18 & $ -1.56$ &   0.37 &   0.0010 &    0.78   & $   46.8$ & $    2.50$ & $    2.43$ & \\
G66-22       &   B &   5336 &   4.41 & $ -0.86$ &   0.12 &   0.0033 &    0.64   & $   10.8$ & $    1.12$ & $    1.16$ & \\
G74-32       &   A &   5872 &   4.36 & $ -0.72$ &   0.30 &   0.0060 &    0.86   & $<   3.0$ & $<   1.05$ & $<   1.04$ & \\
G75-31       &   B &   6110 &   4.02 & $ -1.03$ &   0.20 &   0.0025 &    0.87   & $   40.1$ & $    2.42$ & $    2.36$ & \\
G81-02       &   A &   5959 &   4.19 & $ -0.69$ &   0.19 &   0.0054 &    0.86   & $   25.2$ & $    2.09$ & $    2.06$ & \\
G82-05       &   B &   5377 &   4.45 & $ -0.75$ &   0.09 &   0.0040 &    0.68   & $    8.6$ & $    1.04$ & $    1.08$ & \\
G85-13       &   A &   5728 &   4.38 & $ -0.59$ &   0.28 &   0.0079 &    0.86   & $<   2.0$ & $<   0.75$ & $<   0.76$ & \\
G87-13       &   B &   6185 &   4.13 & $ -1.09$ &   0.20 &   0.0022 &    0.86   & $   44.1$ & $    2.52$ & $    2.45$ & \\
G94-49       &   A &   5473 &   4.50 & $ -0.80$ &   0.31 &   0.0051 &    0.76   & $<   3.0$ & $<   0.69$ & $<   0.71$ & \\
G96-20       &   A &   6393 &   4.41 & $ -0.89$ &   0.28 &   0.0040 &    1.00   & $<   2.0$ & $<   1.26$ & $<   1.21$ & \\
G98-53       &   B &   5948 &   4.23 & $ -0.87$ &   0.19 &   0.0035 &    0.81   & $   23.6$ & $    2.04$ & $    2.01$ & \\
G99-21       &   A &   5587 &   4.39 & $ -0.67$ &   0.29 &   0.0068 &    0.79   & $<   3.0$ & $<   0.80$ & $<   0.82$ & \\
G112-43      &   B &   6174 &   4.03 & $ -1.25$ &   0.24 &   0.0016 &    0.84   & $   41.9$ & $    2.48$ & $    2.41$ & \\
G112-44      &   B &   5919 &   4.25 & $ -1.29$ &   0.22 &   0.0014 &    0.73   & $   43.3$ & $    2.31$ & $    2.25$ & \\
G114-42      &   B &   5743 &   4.38 & $ -1.10$ &   0.19 &   0.0021 &    0.72   & $   24.3$ & $    1.88$ & $    1.86$ & \\
G119-64      &   B &   6281 &   4.18 & $ -1.48$ &   0.28 &   0.0010 &    0.82   & $   32.3$ & $    2.41$ & $    2.35$ & \\
G121-12      &   B &   6028 &   4.23 & $ -0.93$ &   0.10 &   0.0027 &    0.82   & $   43.0$ & $    2.40$ & $    2.35$ & \\
G125-13      &   A &   5948 &   4.28 & $ -1.43$ &   0.27 &   0.0011 &    0.73   & $   38.0$ & $    2.26$ & $    2.20$ & \\
G127-26      &   A &   5891 &   4.14 & $ -0.53$ &   0.24 &   0.0085 &    0.91   & $   17.6$ & $    1.87$ & $    1.86$ & \\
G150-40      &   B &   6068 &   4.09 & $ -0.81$ &   0.16 &   0.0040 &    0.88   & $   28.2$ & $    2.22$ & $    2.18$ & \\
G159-50      &   A &   5724 &   4.37 & $ -0.93$ &   0.31 &   0.0038 &    0.76   & $    5.2$ & $    1.16$ & $    1.16$ & \\
G161-73      &   B &   6086 &   4.00 & $ -1.00$ &   0.16 &   0.0026 &    0.87   & $   44.3$ & $    2.46$ & $    2.39$ & \\
G170-56      &   B &   6094 &   4.12 & $ -0.92$ &   0.17 &   0.0031 &    0.86   & $   32.9$ & $    2.31$ & $    2.26$ & \\
G172-61      &   B &   5325 &   4.47 & $ -1.00$ &   0.19 &   0.0026 &    0.63   & $<   4.0$ & $<   0.67$ & $<   0.70$ & SB1 \\
G176-53      &   B &   5623 &   4.48 & $ -1.34$ &   0.18 &   0.0012 &    0.67   & $   24.6$ & $    1.79$ & $    1.77$ & \\
G180-24      &   A &   6104 &   4.21 & $ -1.39$ &   0.33 &   0.0013 &    0.79   & $   32.9$ & $    2.30$ & $    2.24$ & \\
G187-18      &   A &   5707 &   4.39 & $ -0.67$ &   0.26 &   0.0065 &    0.82   & $<   2.0$ & $<   0.73$ & $<   0.74$ & \\
G188-22      &   A &   6074 &   4.18 & $ -1.32$ &   0.35 &   0.0017 &    0.79   & $   35.2$ & $    2.32$ & $    2.26$ & \\
G192-43      &   B &   6270 &   4.29 & $ -1.34$ &   0.26 &   0.0014 &    0.85   & $   28.9$ & $    2.35$ & $    2.29$ & \\
G232-18      &   A &   5659 &   4.48 & $ -0.93$ &   0.32 &   0.0039 &    0.78   & $<   3.0$ & $<   0.87$ & $<   0.86$ & \\
HD3567       &   B &   6151 &   4.02 & $ -1.16$ &   0.21 &   0.0019 &    0.85   & $   38.3$ & $    2.42$ & $    2.36$ & \\
HD22879      &   C &   5859 &   4.25 & $ -0.85$ &   0.31 &   0.0046 &    0.81   & $   10.4$ & $    1.59$ & $    1.57$ & \\
HD25704      &   C &   5968 &   4.26 & $ -0.85$ &   0.24 &   0.0041 &    0.83   & $   21.5$ & $    2.02$ & $    1.98$ & D \\
HD51754      &   A &   5867 &   4.29 & $ -0.58$ &   0.26 &   0.0078 &    0.88   & $    3.8$ & $    1.15$ & $    1.15$ & \\
HD59392      &   B &   6112 &   3.91 & $ -1.60$ &   0.32 &   0.0008 &    0.80   & $   39.2$ & $    2.38$ & $    2.33$ & \\
HD76932      &   C &   5977 &   4.13 & $ -0.87$ &   0.29 &   0.0043 &    0.85   & $   24.9$ & $    2.09$ & $    2.06$ & \\
HD97320      &   C &   6108 &   4.19 & $ -1.17$ &   0.28 &   0.0021 &    0.82   & $   39.7$ & $    2.41$ & $    2.35$ & \\
HD103723     &   B &   6038 &   4.19 & $ -0.80$ &   0.14 &   0.0039 &    0.85   & $   30.4$ & $    2.24$ & $    2.20$ & \\
HD105004     &   B &   5854 &   4.30 & $ -0.82$ &   0.14 &   0.0036 &    0.79   & $   19.3$ & $    1.88$ & $    1.86$ & \\
HD106516     &   C &   6296 &   4.42 & $ -0.68$ &   0.29 &   0.0066 &    1.03   & $<   1.0$ & $<   0.89$ & $<   0.86$ & SB1 \\
HD111980     &   A &   5878 &   3.96 & $ -1.08$ &   0.34 &   0.0028 &    0.82   & $   49.0$ & $    2.35$ & $    2.30$ & SB1 \\
HD113679     &   A &   5772 &   3.99 & $ -0.65$ &   0.32 &   0.0074 &    0.88   & $   34.0$ & $    2.08$ & $    2.07$ & \\
HD114762A    &   C &   5956 &   4.21 & $ -0.70$ &   0.24 &   0.0056 &    0.87   & $   25.5$ & $    2.09$ & $    2.06$ & SB1 \\
HD120559     &   C &   5512 &   4.50 & $ -0.89$ &   0.30 &   0.0041 &    0.74   & $<   2.0$ & $<   0.55$ & $<   0.57$ & \\
HD121004     &   A &   5769 &   4.37 & $ -0.70$ &   0.32 &   0.0065 &    0.84   & $<   3.0$ & $<   0.96$ & $<   0.96$ & \\
HD126681     &   C &   5607 &   4.45 & $ -1.17$ &   0.35 &   0.0023 &    0.71   & $   14.5$ & $    1.52$ & $    1.51$ & \\
HD132475     &   A &   5746 &   3.76 & $ -1.49$ &   0.38 &   0.0012 &    0.78   & $   53.4$ & $    2.28$ & $    2.23$ & \\
HD148816     &   A &   5923 &   4.13 & $ -0.73$ &   0.27 &   0.0056 &    0.86   & $   18.9$ & $    1.92$ & $    1.90$ & \\
HD159482     &   A &   5837 &   4.31 & $ -0.73$ &   0.30 &   0.0060 &    0.83   & $<   2.0$ & $<   0.85$ & $<   0.84$ & D \\
HD160693     &   A &   5814 &   4.27 & $ -0.49$ &   0.19 &   0.0087 &    0.87   & $    6.4$ & $    1.34$ & $    1.35$ & \\
HD163810     &   B &   5601 &   4.56 & $ -1.20$ &   0.21 &   0.0017 &    0.73   & $   22.3$ & $    1.72$ & $    1.71$ & D \\
HD175179     &   C &   5813 &   4.33 & $ -0.65$ &   0.29 &   0.0069 &    0.85   & $<   2.0$ & $<   0.83$ & $<   0.83$ & \\
HD177095     &   A &   5449 &   4.39 & $ -0.74$ &   0.31 &   0.0060 &    0.73   & $<   3.0$ & $<   0.67$ & $<   0.70$ & \\
HD179626     &   A &   5950 &   4.13 & $ -1.04$ &   0.31 &   0.0030 &    0.80   & $   24.1$ & $    2.05$ & $    2.01$ & \\
HD189558     &   C &   5717 &   3.80 & $ -1.12$ &   0.33 &   0.0025 &    0.83   & $   58.9$ & $    2.31$ & $    2.27$ & \\
HD193901     &   B &   5756 &   4.36 & $ -1.09$ &   0.16 &   0.0020 &    0.72   & $   28.1$ & $    1.97$ & $    1.94$ & \\
HD194598     &   B &   6042 &   4.33 & $ -1.09$ &   0.18 &   0.0021 &    0.82   & $   32.9$ & $    2.27$ & $    2.21$ & \\
HD199289     &   C &   5910 &   4.28 & $ -1.04$ &   0.30 &   0.0029 &    0.79   & $   17.9$ & $    1.88$ & $    1.84$ & \\
HD205650     &   C &   5798 &   4.32 & $ -1.17$ &   0.30 &   0.0021 &    0.73   & $   14.8$ & $    1.70$ & $    1.67$ & \\
HD219617     &   B &   5962 &   4.28 & $ -1.45$ &   0.28 &   0.0011 &    0.73   & $   37.3$ & $    2.26$ & $    2.20$ & D \\
HD222766     &   A &   5434 &   4.27 & $ -0.67$ &   0.30 &   0.0069 &    0.72   & $<   2.0$ & $<   0.47$ & $<   0.51$ & \\
HD230409     &   A &   5418 &   4.54 & $ -0.85$ &   0.27 &   0.0043 &    0.74   & $<   2.0$ & $<   0.46$ & $<   0.49$ & \\
HD233511     &   A &   6106 &   4.23 & $ -1.55$ &   0.34 &   0.0010 &    0.76   & $   33.0$ & $    2.30$ & $    2.24$ & \\
HD237822     &   A &   5703 &   4.33 & $ -0.45$ &   0.29 &   0.0110 &    0.90   & $<   2.0$ & $<   0.73$ & $<   0.75$ & \\
HD241253     &   C &   5931 &   4.31 & $ -1.10$ &   0.29 &   0.0025 &    0.79   & $   19.3$ & $    1.93$ & $    1.89$ & \\
HD250792A    &   B &   5589 &   4.47 & $ -1.01$ &   0.24 &   0.0028 &    0.72   & $   15.6$ & $    1.54$ & $    1.54$ & D \\
HD284248     &   B &   6235 &   4.25 & $ -1.57$ &   0.27 &   0.0008 &    0.80   & $   31.5$ & $    2.36$ & $    2.30$ & \\
\noalign{\smallskip}
\hline
\end{longtable}

\begin{list}{}{}
\item[$^{\rm a}$]
Population classification: A, high-alpha halo; B, low-alpha halo; C, thick disk.
\end{list}

\begin{list}{}{}
\item[$^{\rm b}$]
Note on binarity: SB1, single-lined spectroscopic binary; D, double star.
\end{list}

}% end of onltab

\subsection{Stellar mass and heavy-element fraction}
\label{sect:masses}
Stellar masses have been determined from the Y$^2$ (Yonsei -Yale)
evolutionary tracks published by Yi et al. (\cite{yi03}). Their
stellar models include convective core overshoot and helium diffusion, and
tracks are available for a set of total heavy-element mass fraction,
ranging from $Z = 0.00001$ to $Z = 0.08$. The helium mass fraction
is given by $Y = 0.23 + 2.0\,\,Z$, and the tracks are computed for 
\alphafe = 0.0, 0.3 and 0.6. Oxygen is considered
as an alpha-element, whereas carbon and nitrogen are assumed to follow iron.

For a given value of $Z$, the three available \feh\ - \alphafe\ combinations
lead to practically the same mass tracks. For each star, we have therefore
first calculated $Z_{\rm star}$ from the measured \feh\ and \alphafe\ values and 
then determined two masses by interpolating between tracks in the
\logg\ - log\,\teff\ plane for the two adjacent $Z$-values available. 
Interpolation to  $Z_{\rm star}$ then leads to the mass of the
star. The conversion from \feh\ and \alphafe\ to $Z_{\rm star}$ followed
the procedure adopted for the  Y$^2$ tracks and isochrones (Kim et al.
\cite{kim02}, Table 2). It is based on a solar abundance distribution 
(Grevesse \& Noels \cite{grevesse93}) corresponding to $Z_{\odot} = 0.0181$,  
whereas the more recent 3D, non-LTE solar abundance distribution determined
by Asplund et al. (\cite{asplund09}) corresponds to $Z_{\odot} = 0.0134$.
If this smaller value of the solar heavy-element fraction had been adopted,
the stellar $Z$ values would decrease by 25\%, and the derived masses 
would change somewhat, although we expect the effect on the 
relative masses to be small. When stellar evolutionary tracks based on 
the Asplund et al. (\cite{asplund09}) abundance distribution become
available, improved masses can be derived.
The recent Geneva model grid (Mowlavi et al. \cite{mowlavi12})
is based on this distribution, but 
is limited to $Z$ values from 0.006 to 0.04, an interval which covers 
only a small fraction of our stars.

As discussed in Paper III, there is a small systematic offset
in \logg\ between the Y$^2$  isochrones and our 
nearly unevolved  ($\teff < 5600$\,K) main-sequence stars.
The reason for this is unclear, but could be related   
to a change of the mixing length parameter as a function
of metallicity. To avoid this mismatch we have introduced a
correction of $-0.13$\,dex to the \logg\ values of the Y$^2$
models before using the evolutionary tracks to determine stellar
masses.

The derived masses in units of the solar mass are given
in Table \ref{table:one} together with $Z$ values derived from 
\feh\ and \alphafe .  Based on the statistical one-sigma errors given in
Paper I, $\sigma (\teff ) = 30$\,K, $\sigma (\logg ) = 0.05$\,dex,
$\sigma (\feh ) = 0.03$\,dex, and $\sigma (\alphafe ) = 0.02$\,dex,
we estimate that the statistical error of the mass is 
$\sigma (M / M_{\odot}) \simeq 0.02$. The systematic error is on
the same order of size; an offset of +50\,K in the \teff -scale 
would lead to a systematic change of +0.018 in $M / M_{\odot}$.

Mel\'{e}ndez et al. (\cite{melendez10}) also derived stellar masses
from the Y$^2$ models using both \logg\ and absolute magnitudes
(determined from Hipparcos parallaxes) in combination with log\,\teff .
\alphafe\ is assumed to be +0.3 in all stars with $\feh < -1.0$, which
is close to the average value (+0.27) for the  
24 stars in common with our study. 
The mean mass difference (Mel\'{e}ndez -- this paper) is 
$<\!\!M / M_{\odot}\!\!>\, = 0.014 $ with  $\sigma = 0.028$, 
which supports a statistical
error for $M / M_{\odot}$  on  the order of $\pm 0.02$ in
both works. As seen from Fig. \ref{fig:delmass}, the differences 
between the  two sets of derived masses are closely correlated 
with the corresponding differences in \teff . The scatter around
the regression line 
($\Delta M / M_{\odot} = -0.0047 + 0.00036\,\Delta\teff$) is only
$\pm 0.014 M / M_{\odot}$.

\begin{figure}
\resizebox{\hsize}{!}{\includegraphics{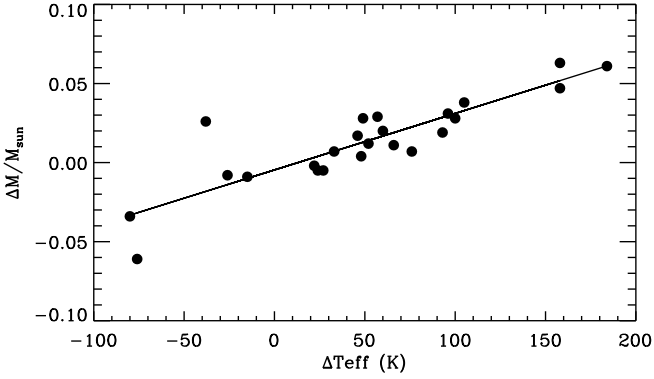}}
\caption{Differences between  stellar masses in  
Mel\'{e}ndez et al. (2012, in preparation) and in this paper
versus the corresponding differences in \teff .}
\label{fig:delmass}
\end{figure}

\section{Lithium versus mass and heavy-element fraction}
\label{sect:Li-relation}

In Fig. \ref{fig:Z-Li.massinterval.all+}, the Li abundances for all stars in 
Table \ref{table:one} are plotted against
the heavy-element mass fraction with different symbols
for the four mass ranges indicated. Disregarding 25 stars with an
upper limit for the Li abundance, it is seen that 
$A$(Li) depends in a systematic way on mass and $Z$. 
$A$(Li) decreases when $Z$ increases, and
for a given $Z$-value, $A$(Li) decreases with decreasing mass.
Corresponding relations have been obtained
by Ryan \& Deliyannis (\cite{ryan98}) and
Boesgaard et al. (\cite{boesgaard05}), who used \teff\ for 
main-sequence stars as a substitute for mass, and by
Mel\'{e}ndez et al. (\cite{melendez10}), who investigated $A$(Li) 
as a function of mass for selected metallicity groups.

\begin{figure}
\resizebox{\hsize}{!}{\includegraphics{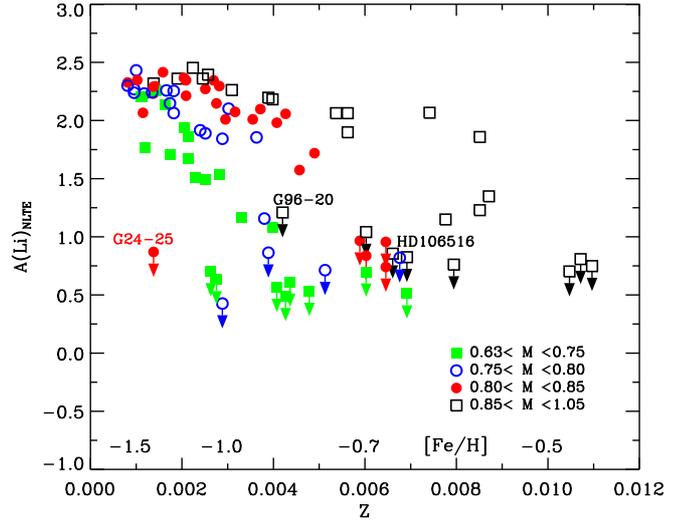}}
\caption{Li abundances as a function of the heavy-element mass fraction, $Z$,
for all stars in Table \ref{table:one}.
Upper limits of $A$(Li) are indicated with a downward directed arrow.
Approximate \feh -values are shown above the x-axis. 
}
\label{fig:Z-Li.massinterval.all+}
\end{figure}

Fig. \ref{fig:Z-Li.massinterval} shows in more
detail how the measured lithium abundances depend on 
$Z$ for the four mass ranges indicated. Stars for which the Li line
could not be detected are excluded from this figure. For a given mass, 
$A$(Li) varies approximately linearly with $Z$ but the slope 
becomes steeper with decreasing stellar mass. As seen from 
Fig. \ref{fig:mass-Li.Zinterval}, the dependence of $A$(Li)
on mass is also close to linear for a given $Z$-value. 
This suggests that $A$(Li) can be fitted with a function 
\begin{eqnarray}
A({\rm Li})_{\rm fit} = a_{0} + a_{1}\,M + a_{2}\,Z + a_{3}\,M\,Z,
\end{eqnarray}
where $a_{0}$, $a_{1}$, $a_{2}$, and $a_{3}$ are constants.
Taking into account the estimated errors, $\sigma \, (A({\rm Li})) = 0.03$,
$\sigma \, (M/M_{\odot}) = 0.02$, and $\sigma \, ({\rm log} \, Z) = 0.04$,
chi-square minimization leads to the expression
\begin{eqnarray}
A({\rm Li})_{\rm fit} = 2.253 + 0.405\,M - 2043\,Z + 2280\,M \,Z,
\end{eqnarray}
where $M$ is in units of the solar mass,
and NLTE Li abundances have been applied. This fit, which is obtained
for 59 stars having $0.7 < M/M_{\odot} < 0.90$ and $Z < 0.006$,
has a reduced chi-square of $\chi ^2 _{\rm red} = 1.12$.
One star, \object{G\,16-20}, within these ranges is
excluded, because it has  a 5.5-sigma deviation in $A$(Li).
If this star is included in the fit, the chi-square
rises to an unacceptably high value of $\chi ^2 _{\rm red} = 1.71$.
\object{G\,16-20} is the most evolved
star in the sample (\logg\ = 3.64), so lithium may have been diluted due to the 
deepening of the convection zone as the star evolved up along the
subgiant branch.

\begin{figure}
\resizebox{\hsize}{!}{\includegraphics{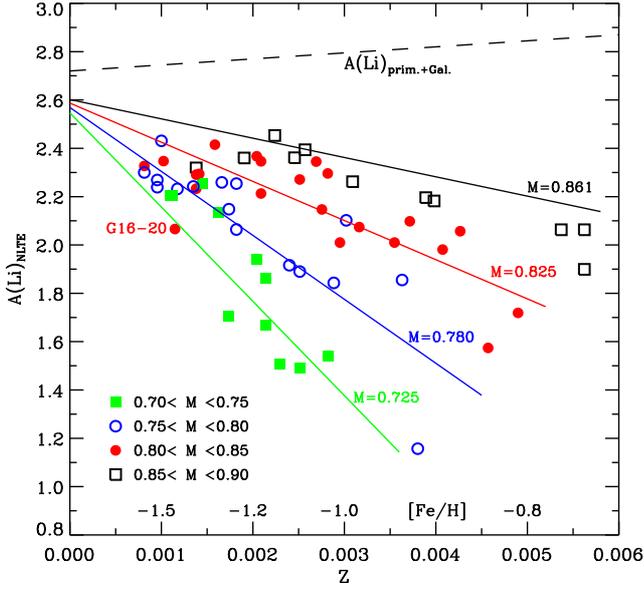}}
\caption{Li abundances as a function of the heavy-element fraction 
for stars with $0.70 < M/M_{\odot} < 0.90$ and $Z < 0.006$. For each
of the four mass intervals, the fit given in Eq. (2) is shown as a straight line
corresponding to the mean mass of the group indicated at the line.
The dashed line shows the sum of the primordial Li abundance
predicted from standard BBN calculations
and $^7$Li made in the Galaxy due to cosmic ray and stellar nucleosynthesis
reactions according to Prantzos (\cite{prantzos07}).}
\label{fig:Z-Li.massinterval}
\end{figure}

\begin{figure}
\resizebox{\hsize}{!}{\includegraphics{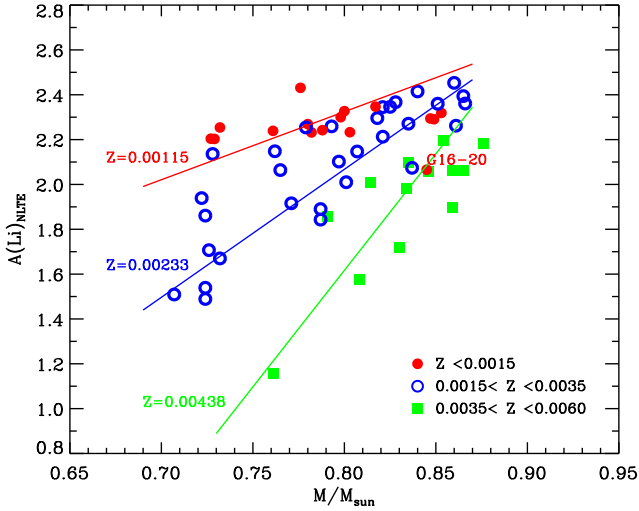}}
\caption{Li abundances as a function of stellar mass
for stars with $0.70 < M/M_{\odot} < 0.90$ and $Z < 0.006$. For each
of the three $Z$-intervals, the fit given in Eq. (2) is shown as a straight line
corresponding to the mean value of $Z$ for the group indicated at the line.}
\label{fig:mass-Li.Zinterval}
\end{figure}

The fit in Eq. (2) is shown as straight lines in  Fig. \ref{fig:Z-Li.massinterval}
for the mean values of $M$ for each of the four mass intervals, and in
Fig. \ref{fig:mass-Li.Zinterval} for the mean values of $Z$ of the three metallicity
groups.  
Some of the dispersion around these lines is due to the ranges
in $M$ and $Z$, respectively, for each group.
The quality of the fit in Eq. 2 can be better seen from
Fig. \ref{fig:Li-Lifit.vs.Z}, where the residuals
are shown as a function of the heavy-element fraction with 
individual error bars given. As seen, the error of 
$\Delta = A({\rm Li}) - A({\rm Li})_{\rm fit}$ increases with
increasing $Z$, which is related to the fact that both the
mass-induced error
\begin{eqnarray}
\sigma (\Delta)_M = (a_1 + a_3 \, Z) \, \sigma(M)
\end{eqnarray}
and the $Z$-induced error
\begin{eqnarray}
\sigma (\Delta)_Z = (a_2 + a_3 \, M) \, \sigma(Z) = (a_2 + a_3 \, M) \,\, {\rm ln} 10 \,\, Z \,\, \sigma ({\rm \, log} Z)
\end{eqnarray}
increase with $Z$.

\begin{figure}
\resizebox{\hsize}{!}{\includegraphics{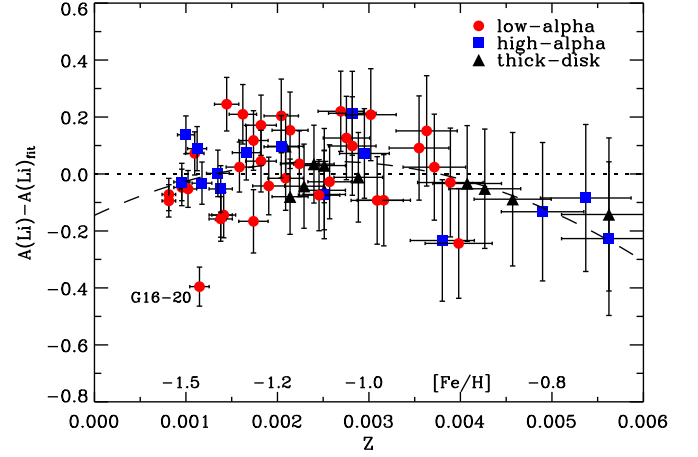}}
\caption{The residuals of the fit given in Eq. (2) as a function
of the heavy-element fraction, $Z$. One-sigma error bars are shown.
The dashed line shows a parabolic fit to the residuals.}
\label{fig:Li-Lifit.vs.Z}
\end{figure}

As seen from Fig. \ref{fig:Li-Lifit.vs.Z}, stars with $0.0015 < Z < 0.003$ 
tend to have positive residuals from the fit in Eq. (2), whereas stars with $Z > 0.004$
have negative residuals. This suggests that there is a small curvature
in the relation between $A$(Li) and $Z$, which is not taken care of in the fit. 
If the residuals are fitted with a parabola as shown with the
dashed line in  Fig. \ref{fig:Li-Lifit.vs.Z},
the reduced chi-square decreases slightly, i.e. from
$\chi ^2 _{\rm red} = 1.12$ to 1.03.

Some of the stars with a non-detectable $\lambda 6707.8$ \LiI\ line 
have an upper limit
of $A$(Li) much below the value predicted from Eq. (2). In a few
cases we may identify the reason for the low lithium abundance.
Three such stars are marked in Fig. \ref{fig:Z-Li.massinterval.all+}.  
\object{G\,96-20} and \object{HD\,106516} have masses around
one solar mass, and may be blue stragglers; they
are significantly bluer than the turn-off of halo and thick-disk
stars with similar metallicities (Carney et al. \cite{carney05}). 
Thus, their low lithium abundance could be due to
merging with a companion star causing Li-depleted gas to be
mixed into the stellar atmosphere. The third star, \object{G\,24-25}, is an  
$s$-process rich star probably due to mass transfer from a former AGB component
(Shu et al. \cite{shu12}), which also brings Li-depleted gas into the
atmosphere. \object{HD\,106516} and \object{G\,24-25} are single-lined 
spectroscopic binaries 
(Latham et al. \cite{latham02}; Ducati et al. \cite{ducati11})
supporting the mass transfer hypothesis as an explanation of the
abnormally low lithium abundance.

As noted in the last column of Table \ref{table:one} several other stars
are single-lined spectroscopic binaries, and five stars are visual 
binaries according to the Hipparcos catalogue
(ESA \cite{esa97}) with components so close (within about one arcsec) that
the component may have affected the observed spectrum.
None of these binary stars stand out with respect to the lithium abundance. 
\object{HD\,219617} is a particularly interesting case. 
According to Takeda \& Takada-Hidai (\cite{takeda11}), the two 
components have nearly the same magnitudes ($V = 8.77$ and 9.08)
and are separated by 0.8\,arcsec. The VLT/UVES spectrum of this star was 
obtained with a 0.8\,arcsec slit under rather poor seeing conditions, and is
thus an unspecified average of the two spectra.

\section{Discussion}
\label{sect:discussion}
According to standard models of F and G main-sequence stars
(e.g. Deliyannis et al. \cite{deliyannis90}),
the depth and thus the temperature at the bottom
of the upper convection zone increase when the stellar mass
decreases. Hence, the destruction of Li by reactions with protons
becomes more effective as the mass decreases, which explains
the dependence of $A$(Li) on stellar mass seen in 
Fig. \ref{fig:mass-Li.Zinterval}. Models
including mixing and diffusion (Pinsonneault et al. \cite{pinsonneault99};
Salaris \& Weiss \cite{salaris01}; Richard et al.  \cite{richard05})
also predict a mass dependence of  $A$(Li) qualitatively similar to what
is seen in Fig. \ref{fig:mass-Li.Zinterval}. 
It remains, however, to be seen if these models can also reproduce
the observed dependence of $A$(Li) on $Z$ for a given mass
(Fig. \ref{fig:Z-Li.massinterval}). 
In any case, the trend of $A$(Li) as a function of mass and $Z$
sets important constraints on stellar modelling.

It should be emphasized that the relation between $A$(Li), $M$, and $Z$
given in Eq. (2) is only valid for main-sequence stars with
$0.7 < M / M_{\odot} < 0.9$ and $0.001 \la Z  < 0.006$
and with a lithium line detected. For stars
more metal-rich than $Z = 0.006$ \footnote{The limit $Z <  0.006$
corresponds to $\feh \la -0.7$ for high-alpha stars and 
$\feh \la -0.6$ for low-alpha stars.}, $A$(Li) shows a large
scatter at a given mass and $Z$, which can be seen from Fig.
\ref{fig:Z-Li.massinterval.all+}. Obviously, other parameters
than $M$ and $Z$ affect the depletion of lithium for these more metal-rich stars.
A similar result has been obtained for F and G main-sequence stars
belonging to the Galactic disk  by Chen et al. (\cite{chen01})
and Lambert \& Reddy (\cite{lambert04}), who
discuss if age and/or initial angular momentum play a role.
The disk stars span a large range in age,
i.e. they have had different times to deplete lithium, and the recent 
work by Baumann et al. (\cite{baumann10}) on lithium abundances of
solar twins shows that $A$(Li) depends quite steeply on age.
The halo and thick-disk stars in our sample have, however,
rather uniform ages, i.e. 10 - 13\,Gyr as shown in Paper III,
and it seems unlikely that differences in age could be the reason
of the large scatter in $A$(Li) for stars with 
$0.006 < Z  < 0.012$.

As seen from  Fig. \ref{fig:Z-Li.massinterval},  
the value of $A({\rm Li})_{Z=0}$, obtained by extrapolating Eq. (2)
to $Z = 0$ for a given mass, is nearly independent of mass.
For the mean mass of the sample, $<\!\!M\!\!> = 0.80\,M_{\odot}$, 
we obtain $A({\rm Li})_{Z=0} = 2.58 \pm 0.04$,
where the (1-sigma) error is determined from the chi-square
analysis used to derive Eq. (2).
The uncertainty of the \teff -scale ($\pm 50$\,K) introduces 
an additional systematic uncertainty of $\pm 0.04$\,dex.
Had we, instead, adopted the parabolic fit shown in 
Fig. \ref{fig:Li-Lifit.vs.Z}, $A({\rm Li})_{Z=0}$ would be lowered by 0.14\,dex,
but this gives too much weight to the high-$Z$ stars in
the extrapolation. An alternative approach is to limit the linear fit
to stars with $Z  < 0.003$, which results in a decrease of 
$A({\rm Li})_{Z=0}$ by 0.08\,dex relative to the value obtained
when using Eq. (2) for the extrapolation. On the other hand,
if the fit is based on the LTE lithium abundances, which would be more in line 
with the small 3D, non-LTE corrections of Sbordone et al. (\cite{sbordone10}), 
$A({\rm Li})_{Z=0}$ increases by approximately 0.08\,dex. Altogether, we estimate 
\begin{eqnarray}
A({\rm Li})_{Z=0} = 2.58 \,\,\pm 0.04_{\rm stat} \,\, \pm 0.10_{\rm syst} .
\end{eqnarray}
Considering the uncertainties, this value is in reasonable agreement  
with the primordial $^7$Li abundance
\begin{eqnarray}
A(^7{\rm Li})_{\rm prim} = 2.72 \pm 0.06
\end{eqnarray}
predicted from standard BBN calculations
(Cyburt et al. \cite{cyburt08}; Coc et al. \cite{coc12})
and the WMAP baryon density (Komatsu et al. \cite{komatsu11}).
 
The derived lithium abundances, therefore,
seem compatible with a scenario where the metal-rich halo stars
were formed out of gas having a lithium abundance equal to that predicted
from standard BBN plus a contribution from Galactic
processes followed by depletion of lithium in the stellar atmospheres
depending approximately linearly on mass and $Z$. The Galactic evolution
of $^7$Li, due to cosmic ray processes
(i.e. $p \, + \, CNO$ and $\alpha \, + \, \alpha$) and Li production
in AGB stars and novae,
has been modelled by Prantzos (\cite{prantzos07}) and is shown  
as a dashed line in Fig. \ref{fig:Z-Li.massinterval}.
As seen, the Galactic contribution to $^7$Li is small compared to
the primordial value and increases linearly with $Z$.
There is also a small contribution to $^6$Li from cosmic ray processes
but since this isotope is much more fragile to proton reactions than 
$^7$Li, the abundance of $^6$Li will be negligible, even for a small
degree of $^7$Li destruction.

\begin{figure}
\resizebox{\hsize}{!}{\includegraphics{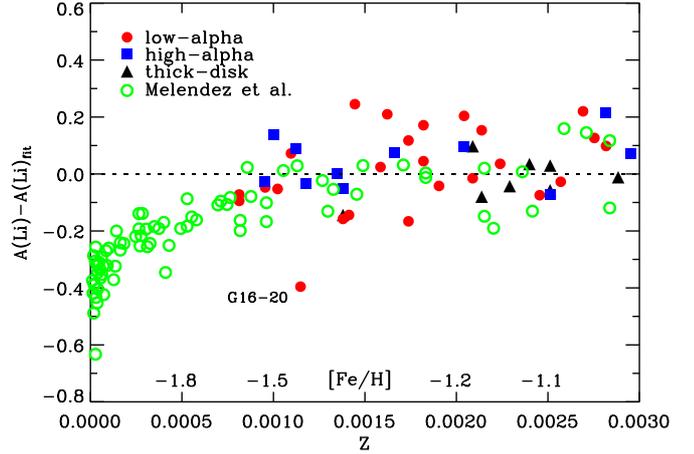}}
\caption{The residuals in Eq. (2) as a function  
of $Z$ with stars from Mel\'{e}ndez et al. (2012, in preparation) included.
}
\label{fig:Li-Lifit.vs.Z.+melendez}
\end{figure}

By extension of this scenario to more metal-poor halo stars, we would
expect that the difference between the lithium abundance measured
for stars on the Spite plateau and that predicted from standard BBN and the 
WMAP baryon density is also due to Li depletion in the stars.
The simple linear relation  between 
$A$(Li), $M$, and $Z$ derived for stars with $0.001 \la Z < 0.006$
is, however, not valid for more metal-poor stars. This can be seen from
Fig. \ref{fig:Li-Lifit.vs.Z.+melendez}, where we have included 
stars from Mel\'{e}ndez et al. (2012, in preparation). 
For $Z \ga 0.001$, the Mel\'{e}ndez et al. stars agree well
with our fit, but at lower metallicities, the Li abundances 
fall below the values expected from Eq. (2). On the Spite plateau
($0.0001 \la Z \la 0.001$), $A({\rm Li}) - A({\rm Li})_{\rm fit}$ is 
nearly independent of mass and $Z$, but at the lowest metallicities
($Z \la 0.0001$) $A({\rm Li}) - A({\rm Li})_{\rm fit}$ 
varies among the stars. Interestingly, the work of Mel\'{e}ndez et al.
(\cite{melendez10}) suggests that this scatter is related to differences
in stellar masses.

As seen from Fig. \ref{fig:Li-Lifit.vs.Z}, the distribution of the
residuals of the fit given in Eq. (2) is nearly the same for
high- and low-alpha halo stars. Given that the high-alpha
stars are likely to have been formed {\em in situ} in the inner
part of the Galaxy, whereas the low-alpha ones were 
accreted from satellite galaxies (Purcell et al. \cite{purcell10};
Zolotov et al. \cite{zolotov09}, \cite{zolotov10}), 
this means that stars belonging to such different
systems were formed with the same Li abundance, and later 
depleted lithium in the same way as a function of mass and $Z$.  
This result is difficult to reconcile with the proposal by
Piau et al. (\cite{piau06}) of astration of lithium in an
early generation of massive, zero-metallicity Pop. III stars.
It would require that the same mass fraction of interstellar gas
has been processed by massive stars in the inner part of the Galaxy
and in satellite galaxies.

The similarity of Li abundances in high- and low-alpha halo stars 
may seem surprising in view of the recent study of beryllium
by Tan \& Zhao (\cite{tan11}). At a given [$\alpha$/H] abundance,
they find the low-alpha stars to be systematically underabundant 
in Be relative to the high-alpha stars. The difference in $A$(Be)
is about 0.2\,dex. Since Be is produced by cosmic ray processes,
i.e. CNO nuclei impinging on interstellar H and He, the explanation
of the lower Be abundance in low-alpha stars may be that the
cosmic ray nucleosynthesis is less efficient in dwarf satellite galaxies than
in the inner part of the Galaxy (Prantzos \cite{prantzos12}).
This would also lead to a 
lower cosmic ray production of lithium, but since the Galactic
part of the initial stellar abundance of lithium is small compared
to the primordial part, it will have only a marginal effect on 
the measured stellar lithium abundances. 

According to Paper III, the high-alpha halo stars have ages
2 -- 3\,Gyr larger than the low-alpha ones. Thus, the
high-alpha stars have had more time to deplete lithium
than low-alpha stars. As we see no significant difference
in their Li abundances, this suggests that 
Li depletion mainly occurred in an early stage of the stellar
evolution, i.e. either in the pre-main-sequence phase or 
near the zero-age main sequence before the star evolved
up towards the turn-off region. 

\section{Concluding remarks}
\label{sect:conclusions}
In this paper, we have made a precise, homogeneous study of
lithium abundances in a sample of main-sequence stars for 
which a previous study (Paper I) has revealed the existence
of two distinct halo populations with different trends of \alphafe\
for the metallicity range $-1.4 < \feh < -0.7$. The kinematics of
the stars and Galactic formation models
suggest that the low-alpha stars have been
accreted from satellite galaxies, including the progenitor of
$\omega$\,Cen, whereas the high-alpha stars were formed {\em in situ} 
in the disk or bulge and later displaced to the halo by the merging 
dwarf galaxies (Purcell et al. \cite{purcell10}; 
Zolotov et al. \cite{zolotov09}, \cite{zolotov10}).

Precise values of the heavy-element mass fraction, $Z$, have been determined 
from \feh\ and \alphafe , and stellar masses from the 
log\,\teff - \logg\ diagram by interpolating between the Y$^2$ 
(Yi et al. \cite{yi03}) evolutionary tracks. For the large majority
of stars with $0.7 < M/M_{\odot} < 0.9$ and $0.001 \la Z < 0.006$
(corresponding to $-1.5 \la \feh \la -0.7$), the lithium abundance
depends quasi-linearly on mass and $Z$ as given in Eq. (2).
It remains to be seen if
stellar models dealing with lithium destruction and depletion
can reproduce this derived relation. Furthermore, it is
an open question why a few stars have a much lower lithium abundance
than predicted by Eq. (2).  

Extrapolating the derived relation between $A$(Li), $M$, and $Z$
to $Z = 0$ yields a lithium abundance close to the value
predicted from standard BBN calculations and the WMAP baryon density.
It is, therefore, tempting to suggest that halo
stars with metallicities $0.001 \la Z < 0.006$ and masses
$0.7 < M/M_{\odot} < 0.9$ formed with a lithium abundance
close to the primordial value and that lithium in their atmospheres
has later been depleted with an almost linear dependence on mass and $Z$.
This suggests that the solution of the `lithium problem' is to
be obtained by getting a better understanding of how lithium
in the atmospheres of metal-poor F and G main-sequence stars
is depleted by diffusion and mixing processes.
It seems unlikely that the solution is to be found in terms
of non-standard BBN or astration of lithium in an early
generation of massive stars.

For stars more metal-poor than $Z \simeq 0.001$, the
lithium abundance is lower than predicted by
Eq. (2), and it is puzzling that stars
with $-2.5 < \feh < -1.5$ on the Spite plateau have the same 
Li abundance for masses in the range 
$0.75 < M/M_{\odot} < 0.82$ (Mel\'{e}ndez et al.  \cite{melendez10}, Fig. 5).
For stars with $\feh < -2.5$, the lithium abundance
seems to vary from star to star depending perhaps
on stellar mass (Mel\'{e}ndez et al.  \cite{melendez10}).
High-precision homogeneous values of $A$(Li), $M$, and $Z$ are needed
for a larger sample of stars spanning the whole metallicity range
of the Galactic halo, in order to learn more about
depletion of lithium in metal-poor stars.

As a main result of this work, we find no significant difference in the 
lithium abundance of stars probably formed {\em in situ}  in the inner
parts of the Galaxy and stars accreted from dwarf galaxies.
This underlines the universality of lithium abundances.
Metal-rich halo stars were 
apparently formed with the same initial Li abundance, and in the course of their
evolution the atmospheric Li content has been depleted at a rate 
depending primarily on stellar mass and metallicity.

\begin{acknowledgements}
We thank the referee, Iv\'{a}n Ram\'{i}rez, for very useful
suggestions leading to improvements of this paper and Jorge Mel\'{e}ndez
for sending an updated version of Table 1 in 
Mel\'{e}ndez et al. (\cite{melendez10}).
This publication made use of the SIMBAD database operated
at CDS, Strasbourg, France, and NASA's Astrophysics Data System.
\end{acknowledgements}

\Online

\end{document}